\documentclass[10pt,letterpaper,twocolumn]{article} 

\usepackage{ol2}
\usepackage[draft]{hyperref}
\usepackage{amsmath}

\begin{document}

\twocolumn[ 

\title{Large phase shift of spatial solitons in lead glass}


\author{Qian Shou, Xiang Zhang, Wei Hu and Qi Guo$^{*}$}

\address{Laboratory of Photonic Information Technology, South China Normal
University, Guangzhou, 510631, China \\$^*$Corresponding author:
guoq@scnu.edu.cn}

\begin{abstract}The phenomenon of the large phase shift of the strongly nonlocal spatial optical soliton was
predicted by Guo $el$ $al$. within the phenomenological framework
[Q. Guo, el al., Phys. Rev. E {\bf 69}, 016602 (2004), but has not
been experimentally confirmed so far. We theoretically and
experimentally investigate the large phase shift
of that propagating in the lead
glass. It is verified that the change of the optical power carried by the optical beam about 10 mW
around the critical power for the soliton can lead to a $\pi$ phase shift, which would be of its potential in the application of all-optical switchings.
\end{abstract}

\ocis{190.5940, 1909.6135, 190.4780}

] 

\noindent The extensively investigated strongly nonlocal spatial
optical solitons (SNSOS) are first predicted by Snyder and Mitchell
\cite{Snyder-Science-1997}. Compared with their local counterparts,
SNSOSs can take on complex forms, such as high order solitons
\cite{Rotschild-PRL-2005,Deng-OL-2007,Rotschild-OL-2006,Deng-JOSA-2007}
and even incoherent solitons
\cite{Krolikowski-PRE-2004,Cohen-PRE-2006,Rotschild-NatPhotonics-2008}.
More importantly, the phase shift of the SNSOS is quite large
\cite{Guo-PRE-2004}. This is an essential attribute of the SNSOSs
but ignored by Snyder and Mitchell \cite{Snyder-Science-1997}.

Assuming that the scalar field of the
monochromatic light is
$$E(x,y,z,t)=A(x,y,z)\exp[-i(\omega t-kz)],$$
$A$ is the paraxial optical beam, $k=\omega n_0/c$, and $n_0$ is the
linear refractive index. $kz$ is a linear phase shift after the
propagating distance $z$ which can be called the geometrical phase
shift, while the argument of the paraxial optical beam, $\arg A$, is
``the additional phase shift'' that will be abbreviated to ``the
phase shift'' in the following. In the linear case,
the increase rate of the phase shift per unit distance is far slower
than that of the geometrical phase shift\cite{Haus}. This is the
reason why the phase shift is not treasured all along. Even in the
nonlinear case, the phase shift per unit distance of the local
soliton was found to be $1/(2kw_{0}^{2})$ \cite{Aitchison-OL-1999},
where $w_{0}$ is the soliton width, which is the same order with the
result for the linear optical beam. Based on the phenomenological
Gaussian response function, Guo $et$ $al$. predicted that the phase
shift rate per unit propagation distance is
$(w_{m}/w_{0})^{2}/(kw_{0}^{2})$ for the SNSOSs\cite{Guo-PRE-2004},
where $w_{m}$ is the width of the response function. The phase shift
is much larger than the local case since the strong nonlocality
means $w_{m}> 10w_{0}$ at least. The phase shift rate in the nematic
liquid crystal, the first-found material with the strongly nonlocal
nonlinearity \cite{Conti-prl-2003-2004},
was found  to be $\pi
^{1/2}(w_{m}/w_{0})/(2kw_{0}^{2})$\cite{Ren-JOA-2008}. Though this
result is slower than that obtained based on the phenomenological
model, it is 10 times faster at least than the results for the local
soliton and the linear beam. It was also pointed out
\cite{Xie-OQE-2004} that $\pi$ phase shift of the signal SNSOS can
be obtained within a very short given distance via adjusting the
pump SNSOS power with the aid of the cross modulation between the
SNSOSs.
Because, however, the additional phase shift of the SNSOS is
completely covered up by the geometrical phase shift during their
propagation, it is somewhat difficult to experimentally demonstrate
the large phase shift of the SNSOS. Someone even doubted that
whether the conclusion of the large phase shift would be
right\cite{Shen-pre-2006}.


In this Letter we theoretically and experimentally investigate the
large phase shift of the SNSOS in lead glass. Based on the principle
of Mach-Zehnder interferometer, we test and verify the linear
modulation of the SNSOS phase by the power of itself. A $\pi$ phase
shift is obtained by changing the soliton power about 10 mw around
the critical power, which demonstrates a high modulation
sensitivity.

The medium we concern is the lead glass with an extremely large range of nonlocality of the thermal self-focusing type nonlinearity\cite{Rotschild-PRL-2005,Alfassi-OL-2007,Rotschild-OL-2006,Rotschild-NatPhotonics-2008}. The propagation
behavior of the light beam in this system is governed by the coupled
equations\cite{Rotschild-PRL-2005,Alfassi-OL-2007}, which are expressed in the cylindrical coordinate system ($R,\phi,Z$) for $Z$-axis symmetric geometry
\begin{subequations}\label{1}
\begin{eqnarray}
&&2ik\frac{\partial A}{\partial Z}+\frac{1}{R}\frac{\partial}{\partial R}(R\frac{\partial A}{\partial
R})+2k^{2}\frac{\Delta
n}{n_{0}}A=0,\label{1a}\\&&\frac{1}{R}\frac{\partial}{\partial
R}(R\frac{\partial T}{\partial R})= -\frac{\alpha}{\kappa}
I(R),\label{1b}
\end{eqnarray}
\end{subequations}
where $\alpha$ and $\kappa$ are respectively the absorption
coefficient and the thermal conductivity, $\Delta n=\beta\Delta T$
with $\beta$ being the thermo-optical coefficient, and $I(R)=|A(R)|^{2}$. We rewrite Eq.~(\ref{1}) in a
dimensionless form:
\begin{subequations}
\begin{eqnarray}\label{2}
&&i\partial_{z} a+\frac{1}{r}\frac{\partial}{\partial
r}(r\frac{\partial a}{\partial r})+Na=0,\label{2a}
\\&& \frac{1}{r}\frac{\partial}{\partial
r}(r\frac{\partial N}{\partial r})=-|a|^{2},\label{2b}
\end{eqnarray}
\end{subequations}
where $r=R/w_{0},z=Z/(2kw_{0}^{2}),a=A/A_{0}$,
$A_{0}^{2}=n_{0}\kappa/(2\alpha\beta k^{2}w_{0}^{4})$ and
$N=2k^{2}w_{0}^{2}\Delta n/n_{0}$.

Following the Snyder's method \cite{Snyder-Science-1997}, the
nonlinear index is expanded and only kept the first two terms of
Taylor series:
\begin{equation}\label{3}
N=N^{(0)}-r^{2}N^{(2)}.
\end {equation}
Assuming the beam of the Gaussian function form
\begin {equation}\label{4}
a =\frac{\sqrt{p_{0}}}{\sqrt{\pi}w(z)}{\rm exp}[i\theta (z)]{\rm
exp}(-\frac{r^{2}}{2w(z)^{2}}),
\end {equation}
where $p_{0}=\int|a(x'-x_{c},y')|^{2}dx'dy'$ is the normalized light
power, we can obtain $N^{(0)}$ by directly integrating Eq.
(\ref{2b}) twice
\begin{equation}\label{5}
N^{(0)}=\frac{p_0}{4\pi}[\Gamma(0,\frac{R_0^2}{w^2(z)})+\ln(\frac{R_0^2}{w^2(z)})+\gamma],
\end{equation}
where $\gamma$ is Euler's constant which equals to $0.5772156649$,
and $R_0$ is the diameter of the cross section of the lead glass.

$\theta$ in Eq. (\ref{4}) is just the phase shift of the beam. We
rewrite it into two terms according to the two terms of the
nonlinear index(in Eq. (\ref{3})):
\begin {equation}\label{6}
\theta=\theta^{(0)}+\theta^{(2)},
\end {equation}
where $\theta^{(0)}=N^{(0)}z$ is the zero-order term of the phase
shift. By the method in Ref. \cite{Guo-PRE-2004}, inserting Eq.
(\ref{3}) and Eq. (\ref{4}) into Eq. (\ref{2b}), the beam width and
the second-order term of the phase shift can be obtained
\begin{subequations}
\begin{eqnarray}\label{7}
&& w(z)=\sigma +(1-\sigma) {\rm cos}(bz) \\&&\theta^{(2)}
=\frac{-2}{2\sigma-1}\{\frac{(1-\sigma)\sin(bz)}{\sigma+(1-\sigma)\cos(bz)},\label{7a}
\nonumber\\&&-\frac{2\sigma}{\sqrt{2\sigma-1}}[\arctan
(\sqrt{(2\sigma-1)}\tan\frac{bz}{2})]\},\label{7b}
\end{eqnarray}
\end{subequations}
where $\sigma=\sqrt{p_{c}/p_{0}}$, $b=2\sqrt{2}/\sigma^{2}$ with
$p_{c}=\pi$ is the critical power for the soliton propagation.

\begin{figure}[ht]
\includegraphics[width=6cm]{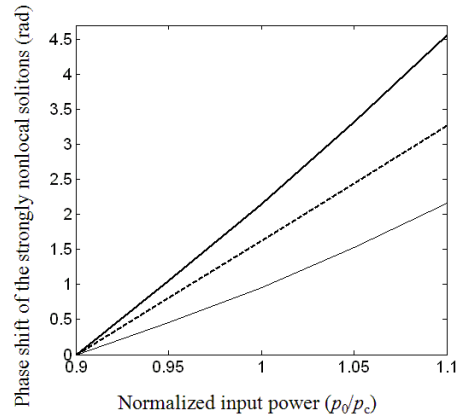}
\centering \caption{Phase shift of SNSOS versus $p_{0}/p_{c}$. The
thick and thin solid lines are respectively the analytical results
in the cases of $w_{0}/R_{0}=1/300$ and $w_{0}/R_{0}=1/60$, which
are the cases in the following experiment. The dashed line is the
numerical result in the case of $w_{0}/R_{0}=1/60$.}\label{Fig.1}
\end{figure}

Based on Eq. (\ref{5}) and Eq. (\ref{7b}), the phase shift is the
function of both the propagation distance $z$ and the power $p_{0}$.
Worthy of note, the zero-order term of the phase shift in Eq.
(\ref{5}) is related to the size of the lead glass. This reveals the
effect of the nonlocality on the phase shift of the SNSOS. In lead
glass the nonlocality is essentially infinite
\cite{Rotschild-PRL-2005} but cut-off by its boundary. Therefore it
could be predicted that larger size glass should owe higher phase
modulation sensitivity. Figure \ref{Fig.1} demonstrates the phase
shift of SNSOS in function of $p_{0}/p_{c}$. We do not provide the
numerical result in the case of $w_{0}/R_{0}=1/300$ because of the
computer source available.

\begin{figure}[ht]
\includegraphics[width=7.5cm]{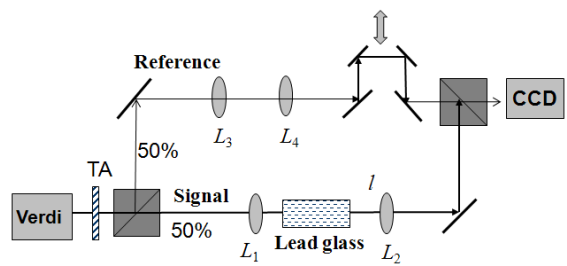}
\centering \caption{Experimental setup. TA is the tunable
attenuator, $L_{1}$, $L_{2}$, $L_{3}$, $L_{4}$ are the
lens.}\label{Fig.2}
\end{figure}

\begin{figure}[ht]
\includegraphics[width=7.5cm]{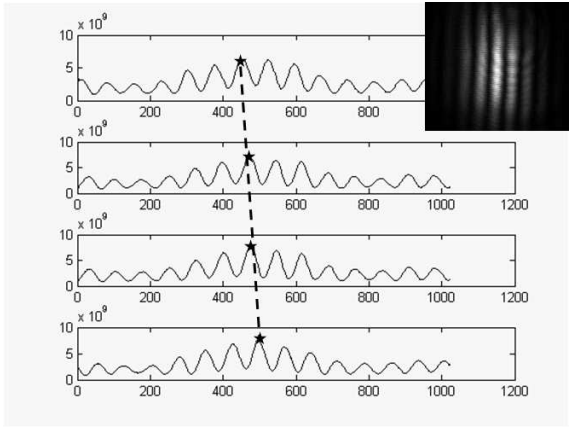}
\centering \caption{Intensity distributions along the direction
perpendicular to the interference fringes through the center of the
fringes. Inset demonstrates the representative interference fringes
between the signal beam and the reference beam.}\label{Fig.3}
\end{figure}

We carry out the large phase shift experiment in cylindrical lead
glass with two diameters of 15 mm and 3 mm but the same length of 60
mm. The heavily-doped glass has a high absorption coefficient
$\alpha$ of 0.07 ${\rm cm}^{-1}$ and a high refractive index $n_{0}$
of $1.9$. The other parameters are the same with those in Ref.
\cite{Rotschild-PRL-2005}. The experimental arrangement is detailed
in Fig. \ref{Fig.2}. A double frequency YAG laser(Verdi 12) with the
wavelength of 532 nm is coupled into a Mach-Zehnder interferometer.
The signal beam on one arm of the interferometer is focused by the
lens $L_{1}$ onto the lead glass with beam width of 50 $\mu$ m. The
output soliton is imaged by $L_{2}$ onto CCD. The other arm contains
a beam telescope, comprised by $L_{3}$ and $L_{4}$, adjusted to give
a collimated, large diameter beam to act as a phase reference.
Considering the difference of the refractive index between the lead
glass and the air, a time delay is used in the reference optical
path to compensate the optical length. The inset of Fig. \ref{Fig.3}
shows representative interference fringes with sharp contrast.

Along with the increase of the SNSOS power, the phase shift of the
SNSOS increases considerably, while the phase of the reference beam
keeps fixed. Therefore the interference fringes move observably. The
stars in Fig. \ref{Fig.3} designate the centers of a tracked fringe.
The equidistant movement of the star indicates a linear modulation
of the SNSOS phase by its power.

\begin{figure}[ht]
\includegraphics[width=8cm]{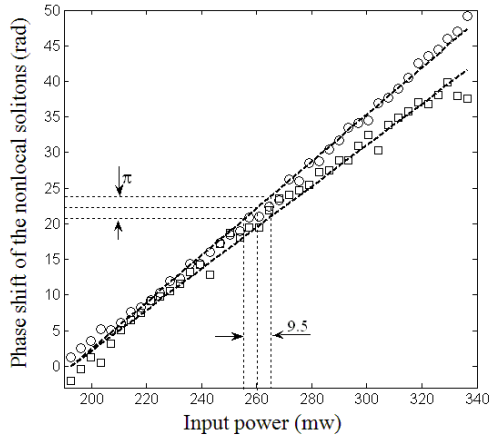}
\centering \caption{Phase shift versus the input power in lead
glasses with different diameters. Circles and squares are
respectively the data obtained in lead glass with diameter of 15 mm
and 3 mm. Dashed lines are the linear fittings with slops of 0.33
and 0.29.}\label{Fig.4}
\end{figure}

Considering the critical power is measured to be 260 mW, we change
the input power from 190 mW to 340 mW by turning the tunable
attenuator TA with power interval of 3.6 mW. Following the procedure
in the treatment of the interference fringes in Fig. \ref{Fig.3}, we
obtain the phase shift in function of the input power in lead
glasses showed in Fig. \ref{Fig.4}. Since the linear fittings of the
experimental data have slops of 0.33 and 0.29, the $\pi$ phase
shifts are modulated by 9.5 mW power change in big glass bar and
10.5 mW power change in small glass bar respectively. In both case,
the power changes are less than $5$\% of the soliton critical power
and therefore the beams almost maintain the form of solitons when
the power is slightly changed. Although the effect of the diameter
of the glass on the phase shift is less than the theoretical
predictions, the modulation sensitivity suggested by the
experimental results is far higher than that predicted by the
theoretical curves in Fig. \ref{Fig.1}. Segev $et$ $al$. were in the
similar situation when they numerically calculated the elliptic
solitons \cite{Rotschild-PRL-2005} and measured the soliton steering
driven by the boundary force \cite{Alfassi-OL-2007}. The
nonlinearity in their experiment is higher and more anisotropic than
the calculated thermal response. The biggest steering data was three
times than the theoretical prediction \cite{Alfassi-OL-2007}. They
presumed an additional mechanism in lead glass, the birefringence
induced by thermal stress, giving rise to an increased $\Delta n$.

In conclusion we investigate the large phase shift of the SNSOS in
lead glass. The experimental result verifies that an output phase
shift of $\pi$ can be linearly modulated by a power change of about
10 mW, which is less than 5\% of the soliton critical power. The
effective producing of $\pi$ phase shift is significant to realize
the treatment and control of the optical signal based on
interference principle.
Additionally the modulation sensitivity is higher in bigger size
glass bar. This is the manifestation that the large phase shift of
the SNSOS stems essentially from the strong nonlocality.


This research was supported by the National Natural Science
Foundation of China (Grant No. 60908003).



\begin{thebibliography}{99}
\bibitem{Snyder-Science-1997} A. W. Snyder  and D. J. Mitchell, Science {\bf 276}, 1538 (1997).
\bibitem{Rotschild-PRL-2005} C. Rotschild, O. Cohen, O. Manela and M. Segev, Phys. Rev. Lett. {\bf 95}, 213904 (2005).
\bibitem{Rotschild-OL-2006} C. Rotschild, M. Segev, Z. Y. Xu, Y. V. Kartashov and L. Torner, Opt. Lett. {\bf 31},
3312 (2006).
\bibitem{Deng-OL-2007} D. M. Deng and Q. Guo, Opt. Lett. {\bf 32}, 3206 (2007).
\bibitem{Deng-JOSA-2007} D. M. Deng, X. Zhao and Q. Guo, J. Opt. Soc. Am. B {\bf 24}, 2537 (2007).
\bibitem{Krolikowski-PRE-2004} W. Kr\'{o}likowski, O. Bang, J. Wyller, Phys. Rev. E {\bf 70}, 036617 (2004).
\bibitem{Cohen-PRE-2006} O. Cohen, H. Buljan, T. Schwartz, J. W. Fleischer and M.
Segev, Phys. Rev. E {\bf 73}, 015601(R) (2006).
\bibitem{Rotschild-NatPhotonics-2008} C. Rotschild, T. Schwartz, O. Cohen and M. Segev, Nat. Photonics {\bf 2}, 371 (2008).
\bibitem{Guo-PRE-2004} Q. Guo, B. Luo, F. Yi, S. Chi, Y. Xie, Phys. Rev. E {\bf 69}, 016602 (2004).
\bibitem{Haus} H. A. Haus, $Waves$ $and$ $fields$ $in$ $optoelectronics$ (Prentice-Hall, 1984).
\bibitem{Aitchison-OL-1999} J. S. Aitchison, A. M. Weiner, Y. Silberberg, M. K. Oliver,  J. L. Jackel,
D. E. Leaird, E. M. Vogel, P. W. E. Smith, Opt. Lett. {\bf 15}, 471
(1999).
\bibitem{Conti-prl-2003-2004}C. Conti, M. Peccianti, and G. Assanto, Phys. Rev. Lett. {\bf91}, 073901
(2003).
\bibitem{Ren-JOA-2008} H. Y. Ren, S. G. Ouyang, Q. Guo, W. Hu and L. G. Cao, J. Opt. A {\bf 10}, 025102 (2008).
\bibitem{Xie-OQE-2004} Y. Q. Xie, Q. Guo, Opt. Quant. Electron. {\bf 36}, 1335 (2004).
\bibitem{Shen-pre-2006}M. Shen, Q. Wang, J. Shi, P. Hou, and Q. Kong, \pre {\bf73}, 056602 (2006).
\bibitem{Alfassi-OL-2007} B. Alfassi, C. Rotschild, O. Manela, M. Segev, Opt. Lett. {\bf 32},
154 (2007).


\end{thebibliography}
\end{document}